\documentclass[amssymb,amsmath,superscriptaddress,footinbib,twocolumn]{revtex4}
\usepackage{natbib,graphicx,subfigure,subeqnarray,fancyhdr,amsmath,multirow,color, mathrsfs}
\begin{document}

\renewcommand{\labelitemi}{$-$}
\newcommand{\change}[1]{{\color{black} #1}}
\newcommand{\Fc}{\mathcal{F}}\newcommand{\Rc}{\mathcal{R}}\newcommand{\dd}{\mathrm{d}}
\newcommand{\ee}{\mathrm{e}}\newcommand{\ci}{\mathrm{i}}\newcommand{\ib}{\mathbf{i}}
\newcommand{\jb}{\mathbf{j}}\newcommand{\kb}{\mathbf{k}}\newcommand{\ab}{\mathbf{a}}
\newcommand{\Fb}{\mathbf{F}}\newcommand{\fb}{\mathbf{f}}\newcommand{\Gb}{\mathbf{G}}
\newcommand{\Mb}{\mathbf{M}Ä}\newcommand{\nb}{\mathbf{n}}\newcommand{\Sb}{\mathbf{S}}
\newcommand{\Sbs}{\mathbf{S^*}}\newcommand{\Rb}{\mathbf{R}}\newcommand{\Sigb}{\boldsymbol{\Sigma}}\newcommand{\sigb}{\boldsymbol{\sigma}}
\newcommand{\Sigbs}{\boldsymbol{\Sigma^*}}\newcommand{\alphab}{\boldsymbol\alpha}
\newcommand{\omegab}{\boldsymbol{\omega}}
\newcommand{\epsb}{\boldsymbol{\epsilon}}
\newcommand{\ub}{\mathbf{u}}
\newcommand{\xib}{\boldsymbol{\xi}}\newcommand{\eb}{\mathbf{e}}\newcommand{\vv}[1]{\underline{#1}}\newcommand{\ev}{\vv{e}}
\newcommand{\rv}{\vv{r}}\newcommand{\TT}[1]{\underline{\underline{#1}}}\newcommand{\omb}{\mathbf{\omega}}
\def\v{\vspace{3cm}}
\newcommand{\Ub}{\mathbf{U}}\newcommand{\xb}{\mathbf{x}}\newcommand{\rb}{\mathbf{r}}
\newcommand{\ssb}{\mathbf{s}}\newcommand{\Xb}{\mathbf{X}}\newcommand{\Pe}{\mbox{Pe}}\newcommand{\Da}{\mbox{Da}\,}
\newcommand{\mean}[1]{\left\langle #1\right\rangle}
\newcommand{\ddp}{[p]^\pm}\newcommand{\taub}{\mbox{\boldmath$\tau$}}\newcommand{\Fr}{\mbox{\textit{Fr}}}
\let\grad\nabla\newcommand{\z}{\zeta}\newcommand{\kk}{\kappa}\newcommand{\tkk}{\tilde{\kappa}}
\newcommand{\e}{\varepsilon}\newcommand{\zb}{\bar{\zeta}}\let\grad\nabla\let\bcdot\cdot
\newcommand{\half}{{\textstyle\frac{1}{2}}}
\newcommand{\textfrac}[2]{{\textstyle\frac{#1}{#2}}}
\newcommand{\LF}[1]{{#1}^{\mathrm{LF}}}\newcommand{\Lap}[1]{{#1}^{\mathrm{L}}}
\newcommand{\ds}{*\!*}\newcommand{\cond}[2]{\frac{\mathrm{D} #1}{\mathrm{D} #2}}
\newcommand{\pard}[2]{\frac{\partial #1}{\partial #2}}\newcommand{\totd}[2]{\frac{\mathrm{d}#1}{\mathrm{d}#2}}
\newcommand{\pardd}[3]{\frac{\partial^2 #1}{\partial #2 \partial #3}}
\newcommand{\Rey}{\mbox{Re}}\newcommand{\Imag}{\mbox{Im}}
\newcommand{\Fpint}{=\!\!\!\!\!\!\!\int}
\newcommand{\txi}{\tilde\xi}\newcommand{\dxi}{\delta\xi}
\newcommand{\tpsi}{\tilde\psi}\newcommand{\dpsi}{\delta\psi}
\makeatletter
\def\sgn{\mathop{\operator@font sgn}}
\makeatother
\allowdisplaybreaks[1]

\title{Geometric pumping in autophoretic channels}
\author{S\'ebastien Michelin}
\email{sebastien.michelin@ladhyx.polytechnique.fr}
\affiliation{LadHyX -- D\'epartement de M\'ecanique, Ecole Polytechnique -- CNRS, 91128 Palaiseau, France}
\author{Thomas D. Montenegro-Johnson}
\affiliation{Department of Applied Mathematics and Theoretical Physics, University of Cambridge, Cambridge CB3 0WA, United Kingdom}
\author{Gabriele De Canio}
\affiliation{Department of Applied Mathematics and Theoretical Physics, University of Cambridge, Cambridge CB3 0WA, United Kingdom}
\author{Nicolas Lobato-Dauzier}
\affiliation{LadHyX -- D\'epartement de M\'ecanique, Ecole Polytechnique -- CNRS, 91128 Palaiseau, France}
\author{Eric Lauga}
%\email{e.lauga@damtp.cam.ac.uk}
\affiliation{Department of Applied Mathematics and Theoretical Physics, University of Cambridge, Cambridge CB3 0WA, United Kingdom}
\date{\today}

\begin{abstract}
Many microfluidic devices use macroscopic pressure
    differentials to overcome viscous friction and generate flows in
    microchannels. In this work, we investigate how the chemical and geometric
    properties of the channel walls can drive a net flow by exploiting the
    autophoretic slip flows induced along active walls by local concentration
    gradients of a solute species. We show that chemical patterning of the wall
    is not required to generate and control a net flux within the channel,
    \change{rather} channel geometry alone is sufficient. Using numerical
    simulations, we determine how geometric characteristics of the wall
    influence channel flow rate, and confirm our results analytically in the
    asymptotic limit of lubrication theory.
    \end{abstract}
\maketitle

\section{Introduction}

Controlled flow manipulation at the micro- or nano-scale is at the heart of
recent developments in microfluidics, including many applications in the field
of biological analysis and screening~\citep{whitesides2006}. Generating and
controlling a flow within the confined environment of a microfluidic channel
requires an external forcing to overcome the viscous stress on the walls. In
synthetic micro and nanofluidic systems, this is usually achieved
\change{either} mechanically, by applying a pressure difference between the
inlet and outlet of the domain, or through
electrokinetics\change{/electroosmosis, where the flow results from an
externally-imposed electric field within the
channel~\citep{squires2005,sia2003microfluidic,ajdari1995,ajdari2000}}.
 
However, many biological systems rely on stresses localized at boundaries in
order to drive flow, rather than on a global macroscopic forcing. For example,
microscopic cilia on the lung epithelium induce a directional flow of mucus
through their coordinated beating, acting as a pump~\citep{sleigh1988}. Similar
microscale flow forcing at the wall also plays an essential role in the early
stages of embryo development~\citep{hirokawa2009} or in the reproduction of
mammals, where cilia-driven flow is responsible for the migration of the ovum
down the female reproductive tract~\citep{halbert1976}.

In a dual {process}, cilia-driven flows play an essential role
in the self-propulsion of micro-organisms such as
\emph{Paramecium}~\citep{brennen1977}; the flow generated by the beating of
cilia anchored on the wall of a moving cell is responsible for its locomotion.
{For both swimming and pumping}, the coordination of neighboring
cilia into metachronal wave patterns {has been proven} essential
to achieving maximum flow rate/swimming speed with a minimum energetic
cost~\citep{gueron1999,michelin2010c,osterman2011,hussong2011,elgeti2013}.

Several attempts have been made to reproduce ciliary pumping in the lab through the
fabrication of artificial actuated
cilia~\citep{dentoonder2008,fahrni2009,babataheri2011,coq2011}. All of them
rely, however, on the application of a global electromagnetic forcing field, and
generating efficient pumping would require the application of phase-shifted
forcing on neighboring cilia~\citep{khaderi2011,khaderi2012}. This {constraint}, as well as
the manufacturing process of microscopic cilia,  poses important challenges to
miniaturization.

Phoretic mechanisms, namely the ability to generate fluid motion near a boundary
under the effect of an external field gradient, represent an alternative route
for both pumping and swimming systems that require no moving parts.
{These mechanisms arise from} the interaction of rigid
boundaries with neutral or ionic solute species in their immediate environment,
and are known to generate the migration of passive particles in external
gradients~\citep{anderson1989}.  Phoretic motion has  recently received
renewed attention in the context of artificial self-propelled systems.
{Such artificial swimmers} generate the {field
gradients required for propulsion} themselves, for example through chemical
reactions catalyzed at their surface, and {thus} do not rely on
any {external} forcing to achieve
propulsion~\citep{paxton2004,howse2007,golestanian2007,palacci2013}.
{These} systems combine two properties: (i) an 
\emph{activity}, i.e.~the ability to release/consume solute species or thermal
energy at their surface, and (ii) a phoretic \emph{mobility}, i.e.~the ability
to create a slip velocity at the boundary from a local tangential gradient
of solute concentration (diffusiophoresis), temperature (thermophoresis) or
electric potential (electrophoresis). Recently, autophoretic systems have
also been considered for generating micro-rotors that rotate without the
application of external torques~\citep{yang2014,yang2015}.

In order to generate the surface gradients necessary to their self-propulsion, autophoretic particles must break spatial symmetry. A  similar
requirement exists for autophoretic pumps. This symmetry-breaking may be
achieved for self-propelled particles following three main routes: (i) chemical
asymmetry, i.e.~patterning the particle surface with active
and passive sites (e.g. Janus
particles)~\citep{golestanian2007,howse2007,theurkauff2012}, (ii) spontaneous
symmetry-breaking resulting from the advection of the field responsible for the
phoretic response by the flow itself~\citep{michelin2013c,izri2014} and (iii)
geometric asymmetry~\citep{shklyaev2014,michelin2015a}.

In this work, we use a combination of theoretical analysis and numerical
simulations to investigate whether, and how, autophoretic mechanisms and
geometric asymmetry can generate a net flow in a microfluidic channel without
imposing any external mechanical forcing, electromagnetic forcing
{or chemical patterning}. For simplicity we focus on the
specific case of diffusiophoresis, where slip velocities are generated at the
wall from tangential gradients in the concentration of a solute released from
one of the channel walls into the fluid. Because of the similarity
between the different phoretic mechanisms, it is expected that the results of
the present contribution may easily be generalized to thermo- or electrophoretic
systems. Specifically, we follow the classical continuum framework of
self-diffusiophoresis~\citep{golestanian2007,julicher2009,sabass2012,michelin2014},
and consider how a left-right asymmetry in the wall shape can generate a net
flow within the channel which hence acts as a microfluidic pump.

The paper is organized as follows. Section~\ref{sec:equations} summarizes this continuum framework for the case of
the flow within an asymmetric channel, and presents the numerical methods used
in this work. Section~\ref{sec:results} shows how the wall geometric
characteristics determine the net flow within the channel. The results are then
confirmed analytically in Sec.~\ref{sec:lubrication} using lubrication theory in
the long-wavelength limit, and conclusions and perspectives are presented in
Sec.~\ref{sec:conclusions}.

\section{Problem formulation}
\label{sec:equations}

\subsection{Diffusiophoretic channel}
We consider a two-dimensional channel of mean height $H$, bounded by a flat
bottom boundary ($y=0$) and a top wall with a periodic non-flat profile,
$y=h(x)$ {(see illustration in Figure~\ref{fig:schema})}. In the channel gap, filled
with a Newtonian fluid of dynamic viscosity $\mu$ and density $\rho$, a solute species of local
concentration $C(\xb)$ with molecular diffusivity $D$ is present and interacts
with the channel walls through a short range potential. When the typical
thickness of this interaction region is much smaller than the other length
scales of the problem (namely the channel gap and the wavelength), the
interaction of the wall with a local solute gradient generates an effective slip
velocity at the wall~\citep{anderson1989,michelin2014}
\begin{equation}\label{eq:mobility}
\ub_\textrm{slip}=M\boldsymbol{\grad}_\parallel C,
\end{equation} 
where $\boldsymbol{\grad}_\parallel=(\mathbf{1}-\nb\nb)\cdot\boldsymbol{\grad}$
is the tangential component of the gradient to the surface of local normal $\nb$
and $M$, the phoretic mobility, is a property of the solute-wall interaction
{which} may be positive or negative depending on the repulsive
or attractive nature of that interaction~\citep{anderson1989}. The chemical
properties of the channel walls are also characterized by a chemical activity,
i.e.~the ability to create or consume the solute species. Here we consider a simple
fixed-flux model, for which the activity of the wall is given by a fixed flux of
solute per unit area $A$, counted positively (resp.~negatively) when solute is
released (resp.~absorbed)
\begin{equation}\label{eq:activity}
D\,\nb\cdot\boldsymbol{\grad} C=-A.
\end{equation} 

In the case of  self-diffusiophoretic propulsion, locomotion
{is often achieved through} inhomogeneity in the chemical
treatment of the particle~\citep{paxton2004,howse2007,golestanian2007}. Recent  work has  shown that geometric asymmetry of chemically-homogeneous
particle alone {is in fact} sufficient to ensure
locomotion~\citep{shklyaev2014,michelin2015a}. Here we investigate a similar
question, namely the possibility of obtaining a net flow from
chemically-homogeneous channel walls using shape asymmetry. We thus assume  that the top corrugated wall has homogeneous mobility $M$ and activity $A$.
To ensure the existence of a steady state solution, the concentration of the
solute {on the bottom wall} is assumed to be fixed
($C=C_0$). {Consequently, the fluid velocity on that wall satisfies  the no-slip boundary condition.} By studying the relative concentration of solute
to {that on the bottom wall, we can assume without loss of
generality that $C_0=0$.}

The phoretic slip velocity generated at the top wall by the wall-solute
interaction drives a flow within the channel. When viscous effects dominate
inertia (namely, when the Reynolds number $\Rey=\rho \mathcal{U}H/\mu$ is small,
with $\mathcal{U}=\change{|AM|}/D$ the typical phoretic velocity), the flow satisfies
{the incompressible Stokes } equations
\begin{equation}\label{eq:stokes}
\mu\grad^2\ub=\boldsymbol{\grad} p,\qquad \boldsymbol{\nabla}\cdot \ub=0,
\end{equation}
for the velocity and pressure fields, $\ub$ and $p$ respectively. 
Solute molecules diffuse within the channel, and in general can also be advected
by the phoretic flows. However, when diffusive effects dominate
(i.e.~when the P\'eclet number, $\Pe=\mathcal{U}H/D$, is small),
the solute dynamics is completely decoupled from the flow, and the solute
concentration satisfies Laplace's equation
\begin{equation}\label{eq:laplace}
\nabla^2 C=0.
\end{equation}

Equations~\eqref{eq:stokes}--\eqref{eq:laplace}, together with the boundary
conditions Eqs.~\eqref{eq:mobility}--\eqref{eq:activity} applied at the top wall
and the inert boundary conditions $C=0$ and $\ub=\mathbf{0}$ at the bottom wall,
form a closed set of equations that can be solved successively for the solute
concentration $C$ and velocity field $\change{\ub = (u,v)}$. From these results, the net flow
rate within the channel, $Q$, can be computed as
\begin{equation}\label{eq:flux}
Q=\int_0^{h(x)}u(x,y)\dd y,
\end{equation}
{which} is independent of $x$ because the flow is incompressible.

\subsection{Asymmetric channel}
\begin{figure}[tb]
\begin{center}
\includegraphics{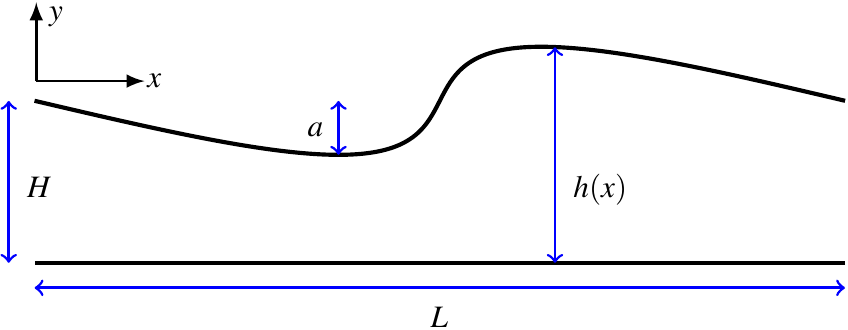}
\caption{Asymmetric phoretic channel. The top wall is characterized by 
constant chemical activity $A$ and mobility $M$. The  {bottom wall}
maintains a fixed concentration and thus flow satisfies  no-slip there. The
example shown here corresponds to $\gamma=\pi/4$, $a/H=1/2$ and $L/H=2\pi$ with
the asymmetric shape of the wall given in Eq.~\eqref{eq:shape}.}
\label{fig:schema}
\end{center}
\end{figure}

The shape of the channel is a periodic function of $x$ characterized by a
wavenumber $k=2\pi/L$. A sinusoidal wall will generate a perfectly
left-right-symmetric concentration distribution and flow pattern, leading to no
net flow along the channel. In the following, we focus on a subset of asymmetric
wall shapes, essentially smoothed ratchets, that are formally obtained by
mathematically shearing the symmetric sinusoidal profile. The top wall is
described  in parametric form by
\begin{equation}\label{eq:shape}
x(s)=s-\frac{\gamma}{k} \sin ks,\qquad y(s)=H+a\sin ks,
\end{equation}
 the non-dimensional \change{asymmetry} parameter, $\gamma$, determines
the asymmetry of the profile, and $H$ and $a$ are the mean channel width and the
amplitude of the width fluctuations, respectively (see Figure~\ref{fig:schema}).

{Hereafter}, the problem is non-dimensionalized using $1/k$ as
characteristic length, $\mathcal{U}$ as characteristic velocity, and $\change{|A|}/Dk$
as characteristic concentration fluctuation. \change{While  $A$ and $M$ are both signed quantities,  after nondimensionalisation we
focus  on the case $A=M=1$;  changing the sign of either $A$ or
$M$ simply reverses the slip velocity forcing and flow rate without changing its
magnitude.} The problem is now completely {specified} by three
geometrical quantities, namely the non-dimensional mean gap, $H^*=kH$, the
corrugation amplitude, $a^*=ka$,  and the \change{asymmetry} parameter, $\gamma$.

\subsection{Numerical method}
Equation~\eqref{eq:stokes}  for the solute concentration and  Eq.~\eqref{eq:laplace} for the  flow and pressure fields are solved numerically using a boundary integral approach with periodic Green's functions. 

We denote $\Omega$ the fluid domain in a period of the channel gap,
$\partial\Omega$ its lower and upper boundaries (the inert and active walls),
and $\nb$ the unit normal vector pointing into the fluid domain. 
The two-dimensional periodic Green's function for Laplace's equation, Eq.~\eqref{eq:laplace}, is given by 
\begin{align}
\Phi(x,y;\xi,\eta) =&\ \frac{1}{4\pi}\sum_{\change{n=}-\infty}^{\infty} \ln\left[(x -
\xi + 2n\pi)^2 + (y - \eta)^2 \right] \nonumber \\ 
=&\ \frac{1}{4\pi}\ln[2(\cosh(y-\eta) - \cos(x-\xi))].
\label{eq:BIE_diffusion}
\end{align}
Assuming that
the channel walls are smooth, the concentration at a point $(x,y)$ on one of the
walls can then be computed {using the} boundary
integral formulation~\citep{banerjee1981boundary}
\begin{align}
\frac{1}{2}C(x,y) = &\int_{\partial\Omega}\left[C(\xi,\eta)\frac{\partial}{\partial
n}(\Phi(x,y;\xi,\eta)) \right.\nonumber\\
&-\left.\Phi(x,y;\xi,\eta)\frac{\partial}{\partial
n}(C(\xi,\eta))\right]\mathrm{d}s(\xi,\eta).
\end{align}
The upper and lower boundaries of the domain are discretized into $200$ straight-line segments, and $C$ and $\mathrm{d}C/\mathrm{d}n$ are assumed constant over each element. For elements on the bottom (resp.~top) boundary, $C = 0$
({resp.~$\mathrm{d}C/\mathrm{d}n=-1$)} is enforced at the
midpoint of each segment. This reduces the boundary integral equation
\eqref{eq:BIE_diffusion} to a dense matrix system for the solution vector
containing the unknown $\mathrm{d}C/\mathrm{d}n$ on the lower boundary and $C$ on the upper boundary. The free-space (i.e.~singular) component of the Green's function is isolated and integrated analytically, and all non-singular element integrals are computed
with a $16$-point Gaussian quadrature. In order to compute the fluid flow and
flow rate in the channel, only the boundary concentration of solute (and not its
bulk distribution) is needed, which makes the boundary element method
particularly suitable for this problem. {The numerical code was
validated against analytical solutions for diffusion in a channel with
nontrivial boundary conditions and domain geometry, achieving a relative error
of at worst $0.004\%$.}

For Stokes flow, the {dimensionless} boundary integral equation
for boundary force density, $\mathbf{f}$, is given by
\begin{align}
    u_j(\mathbf{x}) = \frac{1}{2\pi} &\int_{\change{\partial\Omega}}
[S_{ij}(\mathbf{x}-\boldsymbol{\xi})f_i(\boldsymbol{\xi}) 
\nonumber \\
&- T_{ijk}(\mathbf{x}-\boldsymbol{\xi})u_i(\boldsymbol{\xi}) 
n_k(\boldsymbol{\xi})]\mathrm{d}s(\boldsymbol{\xi}).
\label{eq:BIE_stokes}
\end{align}
For $\hat{\mathbf{x}} = \mathbf{x} - \boldsymbol{\xi}$ and $r =
|\hat{\mathbf{x}}|$, the two-dimensional, $2\pi$-periodic Green's functions for
Stokes flow are 
\begin{equation}
\mathbf{S} = \sum_{n = -\infty}^{\infty} \mathbf{I}\ln r_n -
\frac{\hat{\mathbf{x}}_n\hat{\mathbf{x}}_n}{r_n^2}, \quad
\mathbf{T} = \sum_{n = -\infty}^{\infty}
4\frac{\hat{\mathbf{x}}_n\hat{\mathbf{x}}_n\hat{\mathbf{x}}_n}{r_n^4},
\end{equation}
where $\hat{\mathbf{x}}_n = \left(x - \xi + 2n\pi,y - \eta\right)$.
These functions may be expressed in the closed form 
\begin{equation}
\begin{aligned}
S_{xx} &= K + \hat{y}\partial_{\hat{y}}K - 1, \\
S_{yy} &= K - \hat{y}\partial_{\hat{y}}K, \\
S_{xy} &= -\hat{y}\partial_{\hat{x}}K = S_{yx},\\ 
T_{xxx} &= 2\partial_{\hat{x}}(2K + \hat{y}\partial_{\hat{y}}K),\\
\end{aligned} \qquad
\begin{aligned}
T_{xxy} &= 2\partial_{\hat{y}}(\hat{y}\partial_{\hat{y}}K),\\
T_{xyy} &= -2\hat{y}\partial_{\hat{x}\hat{y}}K,\\
T_{yyy} &= 2(\partial_{\hat{y}}K - \hat{y}\partial_{\hat{y}\hat{y}}K),\\
T_{ijk} &= T_{kij} = T_{jki},
\end{aligned}
\end{equation}
for $K = \frac{1}{2}\ln\left[2\cosh(\hat{y}) - 2\cos(\hat{x}) \right]$. The
computational procedure for discretizing the domain boundary is identical to
that used for the diffusion equation \eqref{eq:BIE_diffusion}. \change{Constant force elements are assumed, singular integrals have the singularity removed and computed analytically, and non-singular integrals are computed with 16- point Gaussian quadrature.}The implementation is based upon the authors' previously published
work {on the optimal swimming of a
sheet}~\citep{PhysRevE.89.060701}, with the addition of Tikhonov regularisation
to improve matrix conditioning.

% here
\section{Results}
\label{sec:results}
\subsection{The role of asymmetry}
When inertia and solute advection are negligible, the Laplace problem for the
solute concentration is linear and decouples from the Stokes flow problem, which
is also linear. Breaking the left-right symmetry is thus required in order to create a net flow within the channel, in the same way that
symmetry breaking is required  {to achieve} self-propulsion of
autophoretic particles. If the chemical properties of the walls are homogeneous,
this asymmetry can only arise from geometry and therefore, purely symmetric
profiles such as sinusoidal upper-wall shapes will {yield} zero net
flux.

\begin{figure}[tb]
\begin{center}
\includegraphics{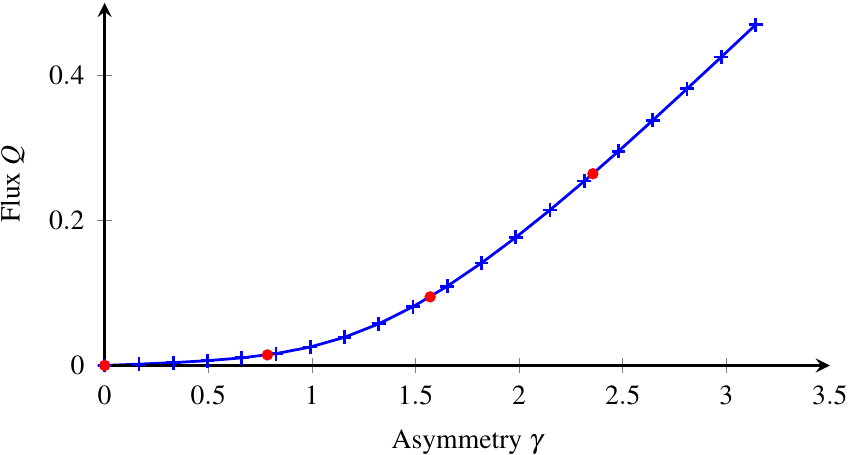}
\caption{Dependence of the net flow rate through the channel, $Q$,  with the
left-right asymmetry of the top wall, $\gamma$, in the cases  $a/H=1/2$, $L/H=2\pi$ and
 {$0 \leq \gamma \leq \pi$}. The four red dots correspond to the
different panels illustrated in Figure~\ref{fig:streamlines}.}
\label{fig:Q_vs_shear}
\end{center}
\end{figure}

In order to analyze the effect of asymmetry, we {first} investigate the
effect of the \change{asymmetry} parameter, $\gamma$, on the flow rate. The numerical results are shown in Figure~\ref{fig:Q_vs_shear}. At $\gamma=0$, we recover zero-net flux, as expected. For asymmetric shapes, we obtain that the flow rate within the channel increases monotonically with $\gamma$.

\begin{figure}[tb]
%\begin{center}
\flushleft
\includegraphics{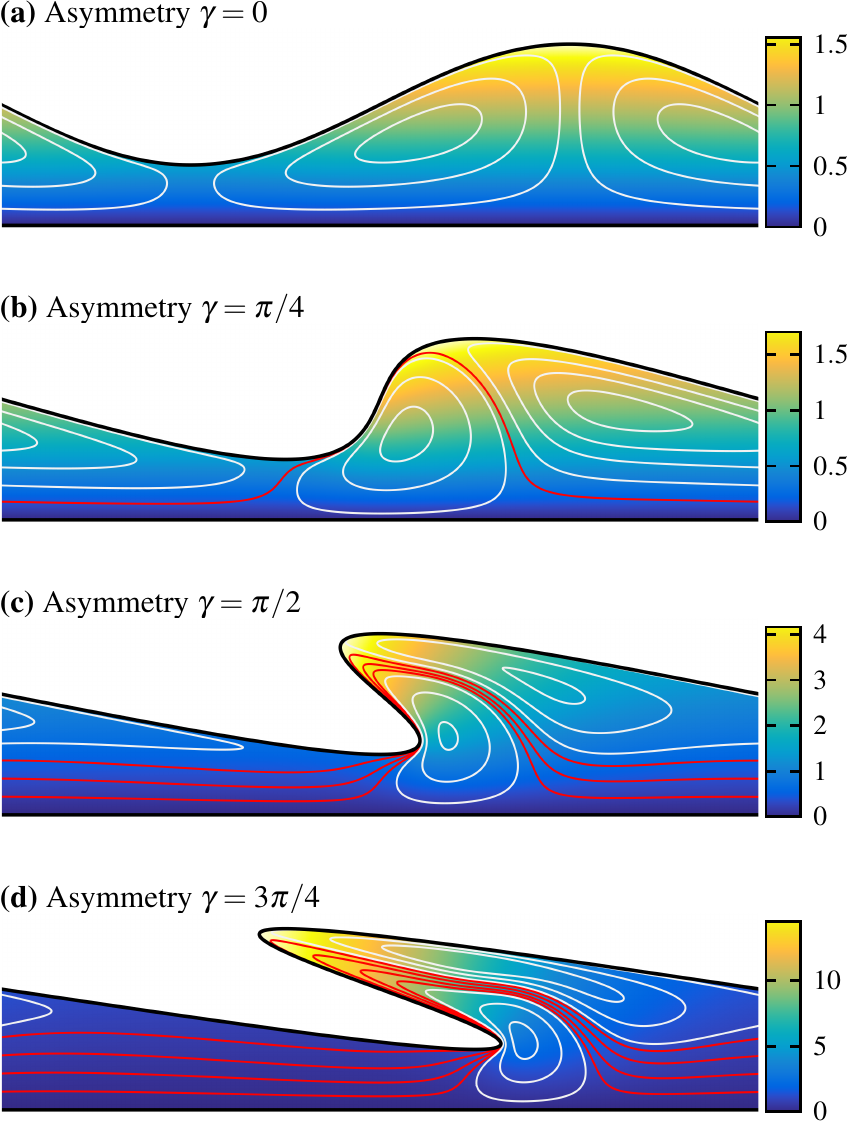}
\caption{Solute concentration (colors) and streamlines (lines) within the channel
with increasing \change{asymmetry} in the case $a/H=1/2$ and $L/H=2\pi$. The four panels above correspond to the four red dots in Figure~\ref{fig:Q_vs_shear}. Red 
(resp.~white) streamlines correspond to traversing (resp.~recirculating) flow regions.}
\label{fig:streamlines}
%\end{center}
\end{figure}

To {gain} insight into the origin of this flow rate, we illustrate in Figure~\ref{fig:streamlines}   the dependence 
of the solute concentration distribution and streamlines with $\gamma$.  For a strictly symmetric profile, $\gamma=0$, a  flow is induced in the channel but one with no net flux. Indeed, a vertical
solute concentration gradient is created between the two walls due to the
fixed-flux emission of solute at the upper wall and the constant concentration
imposed at the lower wall, where the solute is  consumed. The upper
wall is not horizontal, and regions of the upper wall located the furthest from
the bottom wall are exposed to higher concentration than regions corresponding
to the narrowest channel width. This {implies} the existence of
tangential solute gradients along the wall and, hence, of a slip velocity that
drives a flow within the channel. Due to the  symmetry of the channel, the flow
organizes into two counter-rotating flow cells leading to zero fluid transport
across the channel. For positive activity and mobility of the upper wall, the
flow is directed away from the active wall (i.e.~downward) in the regions where
the channel width is greatest, while it is directed \change{toward the active wall (i.e.~upward)} in the
narrowest regions (Figure~\ref{fig:streamlines}a).

When $\gamma\neq 0$, asymmetry is introduced {in two ways}. The
asymmetric upper wall can now be decomposed into a longer backward facing
section and a shorter forward-facing section. The solute gradient on the former
is weaker than on the latter, leading to a stronger left-to-right slip flow
along the forward facing section. {Additionally,} wall asymmetry
increases confinement in the trough along the upper wall leading to higher
solute concentration (the rate of production of solute per unit surface is
fixed). This asymmetry between the wall sections driving the flow within the
channel generates a shape and intensity asymmetry between the two recirculation
regions, and a traversing streamtube appears ({marked by dark red
streamlines in} Figure~\ref{fig:streamlines}). {This streamtube}
corresponds to flow regions that do not recirculate, but are transported along
the channel, being  ``pumped'' by the phoretic
{mechanism}.  This tube follows a pattern along the channel similar to that of
a conveyor belt driven between the two recirculating regions forced by the slip
flow on the wall. Along the shorter forward-facing section of the upper wall, it
is driven by the stronger slip flow that dictates the direction of the net
flow in this case. The tube then separates from the wall where
the slip velocity changes sign, and circumnavigates around the counter-rotating
flow cell driven by the longer wall section.

As $\gamma$ is increased beyond $\gamma\geq 1$, the slope of the
shorter flow-driving wall changes sign, leading to a ``folded'' geometry that
promotes large confinement effects on the solute concentration distribution (see
the difference in color scales in Figure~\ref{fig:streamlines}). This, in turn,
enhances the phoretic slip and net flow rate. For strong asymmetry,
the traversing streamtube is mostly rectilinear and away from the active wall,
except in a narrowing region where it circles around the smaller recirculation
region and is driven by the phoretic slip within the trough on the boundary.

This process does not appear to saturate when $\gamma\gg 1$ for fixed amplitude
$a$. In this limit, the flow domain can be decomposed into two regions: a
{complex}, thin region corresponding to long and thin folds in the wall shape
where very large concentration gradients are established by confinement, and an
outer region where a net unidirectional flow is forced within the channel.
Beyond obvious practical considerations regarding the manufacturing of such
geometries, \change{the} assumptions of the current model would potentially
break down when the \change{asymmetry parameter $\gamma$} becomes too large, as
the phoretic flow become sufficiently intense for advection to be non-negligible
($\Pe\neq 0$). \change{Furthermore, when local concentrations become too large,
it is likely that the model of fixed-flux release would be impacted,
and more detailed reaction kinetics may be required.}

\subsection{Effect of the pattern amplitude on the flow rate}
\begin{figure}[tb]
%\begin{center}
\flushleft
\includegraphics{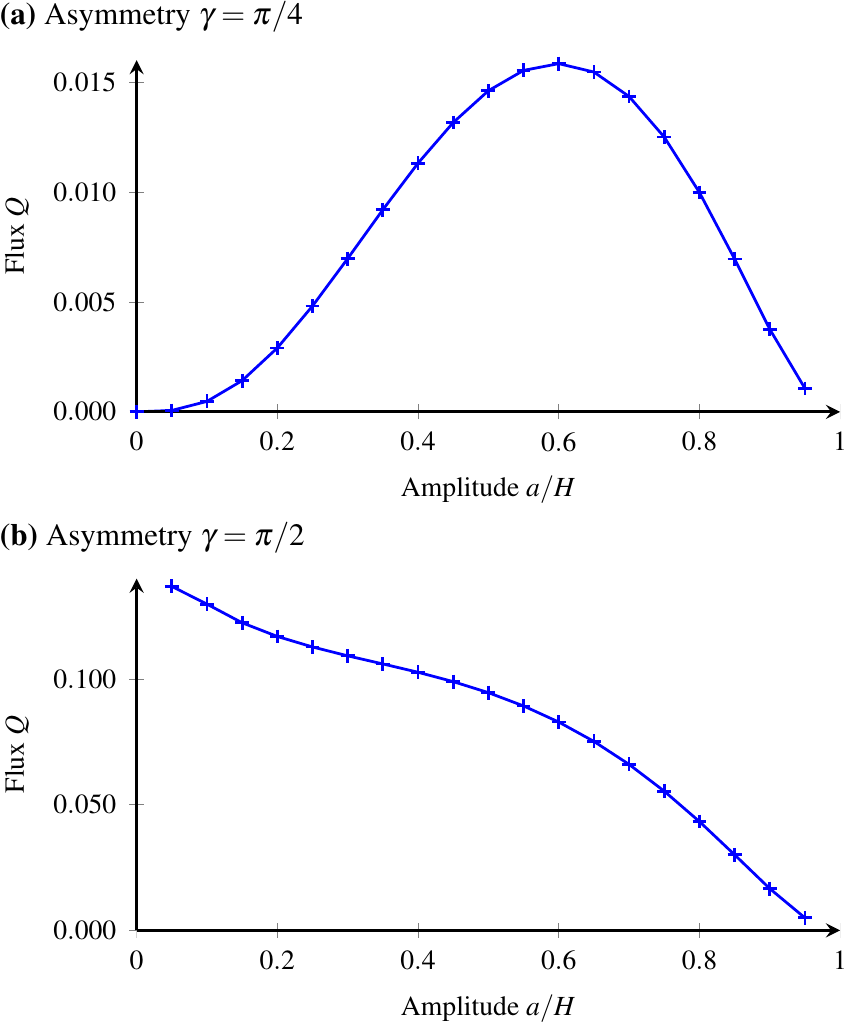}

\caption{The net flow rate through the channel, $Q$, as a function of its
    relative amplitude $a/H$ in the case $L/H=2\pi$. \change{(a) Flux for
        $\gamma=\pi/4$ (as Figure~\ref{fig:streamlines}b), showing behaviour
        representative of weak asymmetry (unfolded) channels. (b) Flux for
        $\gamma=\pi/2$ (as Figure~\ref{fig:streamlines}d), showing distinct
behaviour for strong asymmetry (folded) channels.}}
\label{fig:amplitude}
%\end{center}
\end{figure}

The flow within the channel is effectively driven by the upper wall, while the
no-slip condition on the lower inert wall tends to limit the fluid motion. As a
consequence, it is expected that when the channel gap in the narrowest region
becomes small ($a\approx H$), the net flow rate should be small, as the flow
viscosity will offer maximum resistance there. However, the corrugation
amplitude is an essential element to the pumping performance of the device as it
determines the gradient along the upper wall between the peak and troughs, and
therefore the intensity of the two recirculating regions driving the flow. When
$a\ll H$, it is therefore also expected that the flow rate will become
negligible. This {intuition} is confirmed by our numerical results
 in the case of weak \change{asymmetry} ($\gamma\leq 1$, unfolded
geometry, see Figure~\ref{fig:amplitude}a) for which the net flow rate within the channel displays a maximum at
intermediate amplitude and decreases to zero in both limits $a\ll H$ and
$a\approx H$. In this case, the limit $a\ll H$ corresponds to a flat
upper wall.

The behavior of the system is however quite different when the upper wall is
folded ($\gamma\geq 1$, \change{strong asymmetry}, see Figure~\ref{fig:amplitude}b). In this case, the limit $a\ll H$
 {is not limiting} to a flat wall, but to a surface with
infinitely thin and almost horizontal folds. {Within these
folds,} confinement creates very large solute concentrations and concentration
gradients. As {noted} in the previous section, this limit is
{the singular case} of a flat wall forced periodically by
infinitely large slip velocities in infinitely thin regions. As a consequence,
the net flow rate does not decrease to zero for small amplitude, marking a stark
difference between the folded and unfolded geometries.

\begin{figure}[tb]
\begin{center}
%\flushleft
\includegraphics{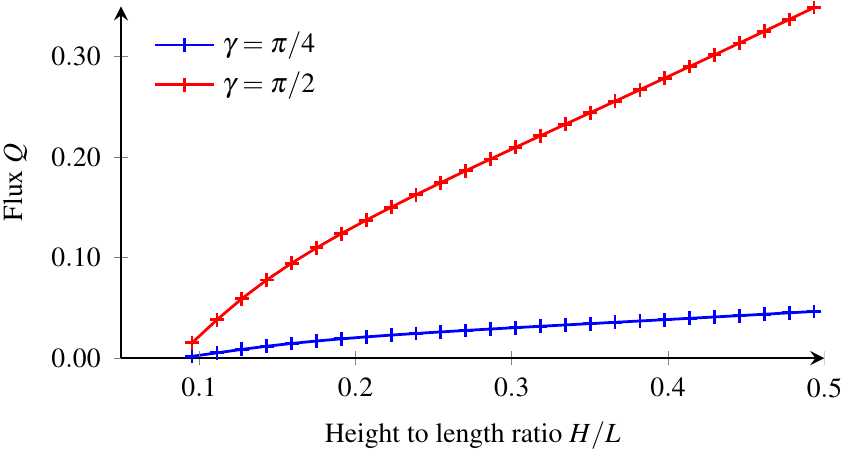}
\caption{{Net flow rate through the channel, $Q$, as a function of
    the channel height $H/L$ for \change{unfolded $\gamma=\pi/4$ and folded
$\gamma=\pi/2$ channels with} fixed relative amplitude, $a/L = 0.08$.}}
\label{fig:height}
\end{center}
\end{figure}

{\subsection{Role of the channel width}
We finally turn to the influence of the third geometric characteristic of the channel, namely its mean width-to-length ratio. The limit of small minimum width ($a\approx H$) was already discussed in the previous section and the flow rate within the channel vanishes in that limit due to the diverging hydrodynamic resistance of the channel.  When $H$ is large compared to both $a$ and $L$, the relative concentration distribution along the top wall is not influence by the location of the passive wall so the tangential concentration gradients and slip velocity become independent of $H$. As a result the net flow rate through the channel varies linearly with the channel height as for a classical Couette (shear) flow. This is confirmed by our numerical results shown in Figure~\ref{fig:height}.}
\section{Long-wavelength prediction}
\label{sec:lubrication}

When the local height of the channel is small in comparison to the typical longitudinal length of the topography, i.e. $|h'(x)|\ll 1$, the problem can be
solved within the {framework of lubrication (long-wavelength) theory}. Defining
$h(x)=\varepsilon f(x)$ with $f(x)=O(1)$ and $\varepsilon\change{=H/L} \ll 1$,
{the method consists in} solving  for the  concentration and velocity fields as  regular series expansions
in $\varepsilon$. On the
upper wall, the normal unit vector pointing into the fluid is now written
\begin{align}
\nb &=\frac{-\eb_y+h'\eb_x}{\sqrt{1+h^{'2}}} 
\nonumber \\
&=-\eb_y+\varepsilon f'\eb_x+\frac{\varepsilon^2f^{'2}}{2}\eb_y-\frac{\varepsilon^3f^{'3}}{2}\eb_x+O(\varepsilon^4).
\end{align}
Defining a rescaled vertical coordinate $y=\varepsilon Y$, the Laplace and
Stokes flow problems are now given by
\begin{subequations}
\begin{align}
\frac{1}{\varepsilon^2}\pard{^2C}{Y^2}+\pard{^2C}{x^2}&=0,\label{eq:lub1}\\
\frac{1}{\varepsilon^2}\pard{^2u}{Y^2}+\pard{^2u}{x^2}&=\pard{p}{x},\label{eq:lub2}\\
\frac{1}{\varepsilon^2}\pard{^2v}{Y^2}+\pard{^2v}{x^2}&=\frac{1}{\varepsilon}\pard{p}{Y},\label{eq:lub3}\\
\frac{1}{\varepsilon}\pard{v}{Y}+\pard{u}{x}&=0,\label{eq:lub4}
\end{align}
\end{subequations}
and the boundary conditions at $y=\varepsilon f(x)$ become
\begin{subequations}
\begin{align}
-1&=-\frac{1}{\varepsilon}\pard{C}{Y}+\varepsilon\left(f'\pard{C}{x}+\frac{f^{'2}}{2}\pard{C}{Y}\right)+O(\varepsilon^3 C),\label{eq:lubbc1}\\
u&=\left(\pard{C}{x}+f'\pard{C}{Y}\right)\left(1-\varepsilon^2f^{'2}\right)+O(\varepsilon^4 C),\label{eq:lubbc2}\\
v&=\varepsilon f'\left(\pard{C}{x}+f'\pard{C}{Y}\right)+O(\varepsilon^3 C).\label{eq:lubbc3}
\end{align}
\end{subequations}
These, together with the conditions $\change{C}=0$ and $u=v=0$ at $\change{Y}=0$, suggest
searching for solutions of the form
\begin{subequations}
\begin{align}
C(x,\change{Y})&=\varepsilon C_1(x,\change{Y})+\varepsilon^3 C_3(x,\change{Y})+O(\varepsilon^5),\label{eq:expc}\\
u(x,\change{Y})&=\varepsilon u_1(x,\change{Y})+\varepsilon^3 u_3(x,\change{Y}) +O(\varepsilon^5),\label{eq:expu}\\
v(x,\change{Y}) & =\varepsilon^2 v_2(x,\change{Y})+O(\varepsilon^4),\label{eq:expv}\\
p(x,\change{Y}) & = \varepsilon^{-1}p_{-1}(x,\change{Y})+\varepsilon p_1(x,\change{Y})+O(\varepsilon^3).\label{eq:expp}
\end{align}
\end{subequations}
The flow rate $Q$ is then {given by}
\begin{equation}
Q=\varepsilon \int_0^{f(x)}u(x,Y)\dd Y=\varepsilon^2 Q_2+\varepsilon^4 Q_4+O(\varepsilon^6),
\end{equation}
with 
\begin{equation}
Q_j=\int_0^fu_{j-1}(x,Y)\dd Y.
\end{equation}
The flow is incompressible and steady, therefore $Q$ and $Q_j$ do not depend on
$x$.

Inserting Eq.~\eqref{eq:expc} into Eqs.~\eqref{eq:lub1} and \eqref{eq:lubbc1}
gives at leading order
\begin{equation}
C_1(x,Y)=Y.\label{eq:c0}
\end{equation}
Eqs.~\eqref{eq:lub2}, \eqref{eq:lubbc2} and \eqref{eq:expu} then provide at $O(\varepsilon)$:
\begin{equation}
u_1(x,Y)=\frac{p_{-1}'}{2}(Y^2-Yf)+\frac{Yf'}{f},
\end{equation}
with $p_{-1}$ the leading-order pressure distribution which is vertically invariant. The function $p_{-1}(x)$ is periodic, therefore
\begin{equation}
Q_2=0 \quad \textrm{and}\quad p_{-1}(x)=-6/f.
\end{equation}
We see that in the lubrication limit, a velocity field is present at
$O(\varepsilon)$, which takes the form of two recirculating regions, but does
not give rise to any net flow through the channel at this order.  After
substitution and {application of} the continuity equation, we
obtain
\begin{subequations}
\begin{align}
u_1(x,Y)&=3\left(\frac{f'}{f^2}\right)Y^2-2\left(\frac{f'}{f}\right)Y,\label{eq:u1}\\
v_2(x,Y)&=\left(\frac{f'}{f}\right)'Y^2-\left(\frac{f'}{f^2}\right)'Y^3.\label{eq:v2}
\end{align}
\end{subequations}

At  next order, the Laplace problem, Eqs.~\eqref{eq:lub1} and \eqref{eq:lubbc1}, together with Eq.~\eqref{eq:c0}, provide
\begin{equation}
C_3(x,Y)=\frac{Y f^{'2}}{2}\cdot
\end{equation}
The horizontal Stokes flow problem now yields,
\begin{align}
u_3(x,Y)&=2\left[\left(\frac{f'}{f}\right)''\left(\frac{Y^3-Yf^2}{3}\right)-\left(\frac{f'}{f^2}\right)''\left(\frac{Y^4-Yf^3}{4}\right)\right]\nonumber\\
        &+\frac{\tilde{p}_1'}{2}(Y^2-\change{Y}f)+\left(ff'f''-\frac{f^{'3}}{2}\right)\frac{Y}{f},
\end{align}
with $\tilde{p}_1(x)$ a function of $x$ only.
\change{Integrating the previous equation in $Y$ finally provides the flow rate at $O(\varepsilon^4)$}
\begin{align}
Q_4&=2\left[ \frac{3f^5}{40}\left( \frac{f'}{f^2}\right)'' - \frac{f^4}{12}\left( \frac{f'}{f}\right)''\right]  -\frac{\tilde p'_1}{12}f^3\nonumber\\
&+ \left(\frac{f^2f'f''}{2}-\frac{f f'^3}{4} \right).
\end{align}
Using the periodicity of $\tilde{p}_1$, $Q_4$ can be computed by dividing the previous equation by $f^3$, taking the spatial average in $x$, and integration by parts.
The result can be rewritten in terms of the original channel height $h(x)$. 
{At leading order} we obtain that the flow  {through} the
channel {is given by} 
\begin{equation}\label{eq:flowratelubric}
Q=\frac{11}{30}\frac{\mean{h^{'3}/h^2}}{\mean{1/h^3}},
\end{equation}
{where $\mean{\cdot}$ is the spatial average over a period.}

We see  two important results: (i) the flow rate is intimately linked to the distribution of local
slope along the wall $h'(x)$, and (ii) shape asymmetry is essential for the
existence of a net flow. Indeed, slip flow along the active wall arises from the
orientation of the wall with a component along the leading order solute
concentration gradient. A non zero $h'(x)$ is therefore sufficient to guarantee
the existence of a local flow but not necessarily of a net flow through the
channel. This separation of scales is clearly visible in the lubrication
expansion: the leading order flow arises from the local channel geometry (i.e.~the fact that the wall is not flat and orthogonal to the leading
order solute gradient).  However, at this order, the net flow is zero because the
flows driven by the forward- and backward-facing walls exactly cancel out. {A net flux} results from an imbalance between these local flows 
which can only be induced by  geometric asymmetry. Left-right
symmetric profiles are  characterized by an even channel height, $h(x)=h(-x)$. Consequently the  function 
$ h^{'3}/h^2$ is  odd and thus  exactly averages to zero, so that $Q_4$ is identically zero (all higher orders are expected to be zero as well).

The result in
Eq.~\eqref{eq:flowratelubric} is a weighted algebraic spatial
average of the slope of the active wall. More precisely, the leading order flow rate {through} the channel,
Eq.~\eqref{eq:flowratelubric}, is the ratio of two integrals;
the numerator is the mean flow forcing due to the  asymmetry of the channel, while the
denominator is the average hydrodynamic resistance of the channel over a
wavelength.

\begin{figure}[tb]
\begin{center}
\includegraphics{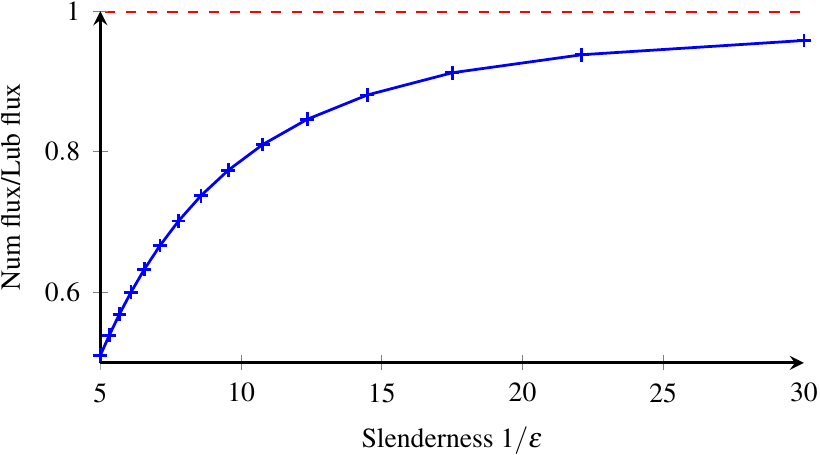}
\caption{Ratio  between the numerical
results and the   lubrication (long-wavelength) theory predictions for  $\gamma=\pi/10$ and $a=H/3$ as a function of an increasing slenderness, $1/\varepsilon$.}
\label{fig:lubrication}
\end{center}
\end{figure}

This leading-order prediction is compared to the full numerical simulations in
Figure~\ref{fig:lubrication}. For a fixed \change{asymmetry parameter} $\gamma$ and relative amplitude
$a/H$, several simulations are performed for increasing $L/H$ (note that as $H$
is reduced, $a$ is reduced in the same amount) and the flow rate through the channel is shown
to converge for large slenderness to the prediction of the lubrication theory.

%BEGIN
Note that the lubrication result, Eq.~\eqref{eq:flowratelubric}, is valid for
any ratio {$H/L\ll1$}, regardless of the relative magnitude of the
mean channel height $H$ and the perturbation amplitude $a$. In the limit of
small wall roughness ($a\ll H$), 
 the hydrodynamic resistance
(the denominator in Eq.~\ref{eq:flowratelubric}) is independent of $a$ at
leading order and simply scales as $1/H^3$, while the phoretic forcing (the numerator in Eq.~\ref{eq:flowratelubric}) scales as $a^3/H^2$. As a consequence, $Q$ scales as $a^3 H$ when $a\ll H$. 

In the opposite limit $a\sim H$, using classical asymptotic expansions to compute the leading order contribution to the integrals in Eq.~\eqref{eq:flowratelubric}~\citep{bender1978}, one can show that the
flow forcing due to the channel's asymmetry (i.e.~the numerator in
Eq.~\ref{eq:flowratelubric}) remains finite and $O(H)$, but that in contrast the hydrodynamic resistance
diverges. More specifically, \change{a standard lubrication calculation} leads at leading order to
\begin{equation}
\mean{\frac{1}{h^3}}=\frac{3\sqrt{2}}{16H^{1/2}(H-a)^{5/2}}\cdot
\end{equation}
 As a consequence, $Q\sim H^{3/2}(H-a)^{5/2}$ when $(H-a)\ll H$. 
 
 Note that since
the limiting factor is {then the channel} width at the narrowest
point, and its impact on the hydrodynamic resistance, it is expected that this
scaling in $(H-a)^{5/2}$ should hold true even when $H$ is not small and could be recovered through a new lubrication expansion limited to the narrow-gap region of the channel~\citep{leal2007}, provided the curvature of the wall in that region remains finite. 
%END

\section{Conclusions}
\label{sec:conclusions}

The active  research in recent years on  autophoretic particles has demonstrated that  fuel-based mechanisms represent a promising route to designing self-propelled  systems that rely only on chemical reactions and the interaction with  the immediate environment to create locomotion. The results presented in this work show that this is {also true for} the dual problem of pumping flow within a micro-channel, and that geometric asymmetry, rather than chemical patterning of the channel walls, is sufficient to create a net flow. Our results  provide insight into the flow dynamics within the channel, and the mechanism leading to the net fluid transport: the breaking of symmetry between two recirculating flow regions driven by  wall slip velocity, and the emergence of a conveyor-belt-like flow within the
channel.

For simplicity, we focused in this paper on a reduced set of wall shapes with one
 active  wall and the other one passive.  
{The numerical methodology and the long-wavelength theory, 
Eq.~\eqref{eq:flowratelubric}, are however valid for any periodic channel geometry in two dimensions.}  Furthermore, these results were obtained within the simplified framework of a  fixed-rate
release of solute by the active wall. Previous studies on autophoretic
self-propelled particles have shown that the exact reaction kinetics, in
particular the dependence of the reaction rate on the local solute
concentration, may significantly impact the system
dynamics~\citep{ebbens2012,michelin2014}. 
We expect for example the  direction of pumping to be  impacted by reaction kinetics, although the basic result showing the 
emergence of a net flow due to geometric asymmetry of the phoretic wall should
remain true. \change{Finally,  our study focused on the particular case of self-diffusiophoresis.  Because of the formal similarities in the equations of the problem, these results can be generalized easily to other phoretic mechanisms such as self-thermophoresis or self-electrophoresis.}

\begin{figure}[t]
\begin{center}
\includegraphics{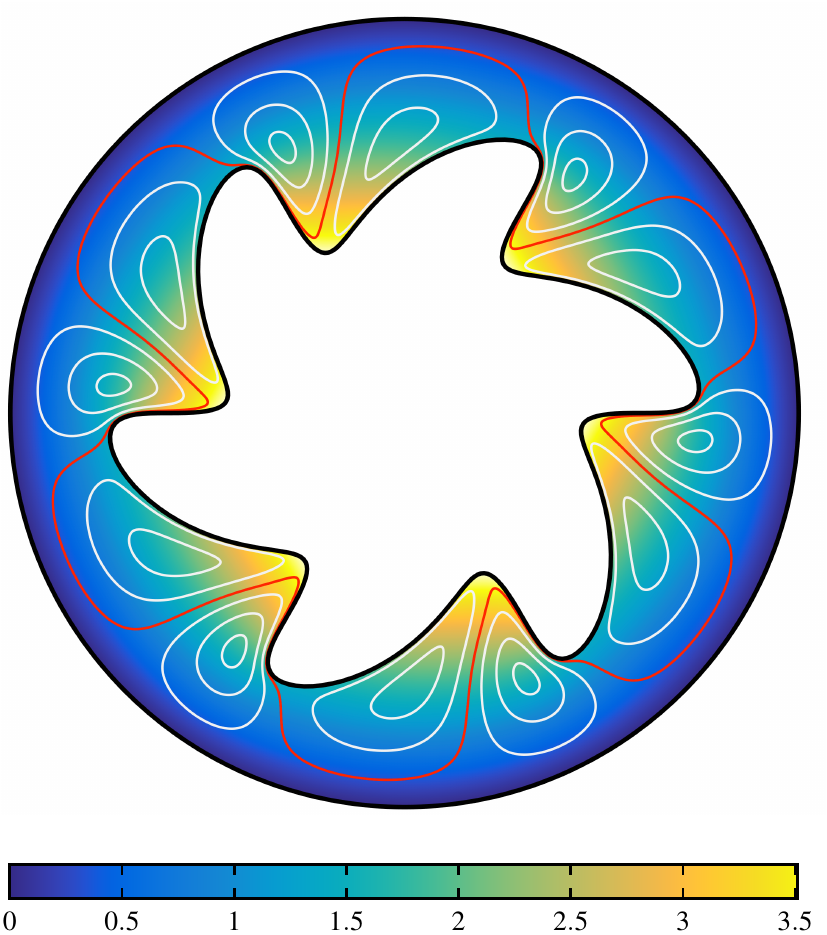}
\caption{Concentration distribution and flow streamlines within an annular  closed-loop channel with a geometrically-asymmetric inner active wall releasing solute with a fixed flux, and a passive circular outer wall with uniform concentration. The recirculating streamlines are shown in white while the traversing streamlines are plotted in red, and correspond to a clockwise-rotating flow.}
\label{fig:ninja}
\end{center}
\end{figure}

Looking forward, the results of this work could be generalized to a larger range of geometries, including closed-loop channels for which pressure-driven flows can not easily be achieved. This is shown in  Figure~\ref{fig:ninja} where we have adapted our numerical approach to compute the   net clockwise flow through a two-dimensional annular channel  driven by the geometric asymmetry of the inner active wall.

\section*{Acknowledgements}
SM acknowledges the support of the French Ministry of Defense (DGA). TDMJ is
supported by a Royal Commission for the Exhibition of 1851 Research Fellowship.
GDC acknowledges support from Associazione Residenti Torrescalla.  This work was
funded in part by a European Union Marie Curie CIG Grant to EL.

\footnotesize{
\providecommand*{\mcitethebibliography}{\thebibliography}
\csname @ifundefined\endcsname{endmcitethebibliography}
{\let\endmcitethebibliography\endthebibliography}{}

%\bibliography{refs_channel.bib} %your .bib file
%\bibliographystyle{rsc} %the RSC's .bst file
}

\end{document}